\documentclass[final,3p,twocolumn]{elsarticle}
\usepackage{amsmath}
\usepackage{graphicx}
\usepackage{amssymb}
\usepackage{times}

\begin{document}

\title{Subsonic Wind Accretion}

\author{Andrei Gruzinov}

\begin{abstract}
The rate of subsonic wind accretion (accretion on a point Newtonian mass moving through uniform gas) is shown to be independent of the wind velocity and equal to the spherical Bondi rate -- for the adiabatic index equal to 5/3. A (very accurate) numerical calculation of the accretion flow, confirming this result, is also presented. 

\end{abstract}

\maketitle

\section{Introduction}
\label{sec:intro}

Bondi's calculation of the spherical accretion \cite{Bondi} is a pearl of theoretical astrophysics. 

Although Bondi is not a priori applicable to real (magnetized, turbulent, and often collisionless) systems, e.g. \cite{Shva, Ress}, stationary hydrodynamics is still the right starting point, because hydrodynamics meaningfully accounts for the stress-energy conservation. 

Here we show that, for adiabatic index equal to 5/3, the subsonic wind accretion problem reduces exactly to the spherical Bondi problem near the accretor. As a result, the mass accretion rate does not depend on the wind velocity. The mass accretion rate is given by the spherical Bondi formula. \footnote{The following calculations are simple, the result is interesting, but the author was unable to find it in the literature -- would be grateful for a reference.}

Theoretical considerations are in \S\S\ref{sec:theo}, \ref{sec:md}, numerical results, confirming the theory, are in \S\ref{sec:num}.

We must acknowledge that both our theory and our numerics have only a provisional status  -- the results should be confirmed by a direct three-dimensional numerical simulation. For all we know, the actual flow may be turbulent, while we assume stationarity.

~

\section{Potential Accretion}
\label{sec:theo}

The wind accretion problem reads: 

Find the gas flow onto a point Newtonian mass $M$. At infinity, the gas is uniform, with the density $\rho_\infty$, the sound speed $c_\infty$, and the velocity $V$. Here and below, the adiabatic index of the gas is $\gamma=5/3$.

Bondi \cite{Bondi} shows that, for $V=0$, the mass accretion rate is 
\begin{equation}\label{eq:mdb}
\dot{M}_B=\frac{\pi(GM)^2\rho_\infty}{c_\infty^3}.
\end{equation}

We will show that this formula is valid for any $V<c_\infty$. We will also derive a variational principle which will allow a (very accurate) numerical calculation of the flow. 

Assume that subsonic flows ($V<c_\infty$) have no shocks. Since the flow at infinity is potential, ${\bf V}=\nabla({\bf V}\cdot {\bf r})$, and the Newtonian gravitational force is potential, the entire accretion flow will be potential, with the velocity field ${\bf v}({\bf r})$ given by the scalar potential $\phi({\bf r})$,
\begin{equation}\label{eq:potf}
{\bf v}=\nabla \phi. 
\end{equation}

The stationary flow equation then follows from the variational principle
\begin{equation}\label{eq:varp}
\delta S=0,
\end{equation}
\begin{equation}\label{eq:act}
S=\int d^3r L^{5/2},
\end{equation}
\begin{equation}\label{eq:lag}
L\equiv 1+\frac{V^2}{3}+\frac{1}{3r}-\frac{(\nabla \phi)^2}{3},
\end{equation}
where $r$ is the spherical radius and the units are 
\begin{equation}\label{eq:uni}
\rho_\infty=c_\infty=2GM=1.
\end{equation}

Indeed, by varying the action, we get the correct continuity equation
\begin{equation}\label{eq:cont}
\nabla \cdot (\rho \nabla \phi)=0,
\end{equation}
where
\begin{equation}\label{eq:den}
\rho = L^{3/2}
\end{equation}
is the correct expression for the density -- by the Bernoulli equation. 

We will use Eqs.(\ref{eq:cont}, \ref{eq:den}, \ref{eq:lag}) in \S\ref{sec:md} to find the mass accretion rate. We will use the variational principle in \S\ref{sec:num} for numerical calculations. 

\section{The Mass Accretion Rate}
\label{sec:md}

To find the mass accretion rate $\dot{M}$ analytically, first define the differential mass accretion rate 
\begin{equation}\label{eq:dacr}
\frac{1}{2\pi \sin \theta}\frac{\partial \dot{M}}{\partial \theta}=-r^2\rho \partial_r \phi,
\end{equation}
where $\theta$ is the polar angle.

Assuming axisymmetry, following Bondi, using Eqs.(\ref{eq:lag}, \ref{eq:den}), we get 
\begin{equation}\label{eq:bone}
\begin{split}
\left( \frac{1}{2\pi \sin \theta}\frac{\partial \dot{M}}{\partial \theta}\right) ^2(r^3\rho^2)^{-1}+3(r^3\rho^2)^{1/3}=\\
1+\left(3+V^2-r^{-2}(\partial _\theta\phi)^2\right)r.
\end{split}
\end{equation}

Now we need to estimate $\partial _\theta\phi$ for small $r$. To leading order,
\begin{equation}\label{eq:lorp}
\phi(r,\theta)=r^{1/2}\phi_0(\theta). 
\end{equation}
Then Eqs.(\ref{eq:lag}, \ref{eq:den}) give, to leading order,
\begin{equation}\label{eq:lord}
\rho(r,\theta)=r^{-3/2}\rho_0(\theta).
\end{equation}
Now the continuity equation (\ref{eq:cont}) reads
\begin{equation}\label{eq:conl}
\frac{d}{d\theta}\left( \sin \theta \rho_0(\theta)\frac{d\phi_0 }{d \theta} \right)=0,
\end{equation}
giving 
\begin{equation}\label{eq:pnd}
\frac{d\phi_0 }{d \theta}=0.
\end{equation}

We see that $\partial _\theta\phi$ is at most $r^{3/2}$ at small $r$ . Eq.(\ref{eq:bone}) then reads, to first {\it two} leading orders, 
\begin{equation}\label{eq:bons}
\begin{split}
\left( \frac{1}{2\pi \sin \theta}\frac{\partial \dot{M}}{\partial \theta} \right) ^2(r^3\rho^2)^{-1}+3(r^3\rho^2)^{1/3}=\\
1+(3+V^2)r.
\end{split}
\end{equation}

On each sphere $r={\rm const}$, there must exist at least one circle $\theta=\theta_0(r)$ where the local mass accretion rate equals the mean accretion rate:
\begin{equation}\label{eq:mean}
\frac{1}{2\pi \sin \theta}\frac{\partial \dot{M}(r,\theta_0(r))}{\partial \theta}=\frac{\dot{M}}{4\pi}.
\end{equation}
Then 
\begin{equation}\label{eq:bons0}
\begin{split}
\left(\frac{\dot{M}}{4\pi} \right) ^2(r^3\rho^2)^{-1}+3(r^3\rho^2)^{1/3}=\\
1+(3+V^2)r.
\end{split}
\end{equation}
Here and below
\begin{equation}\label{eq:rhos}
\rho\equiv \rho(r,\theta_0(r)).
\end{equation}

By the physics of the problem, to account for the assumed ``perfect suction'' of the accretor, we replace the Newtonian potential by the Paczynski-Wiita potential
\begin{equation}\label{eq:pwp}
\frac{1}{r}\rightarrow\frac{1}{r-r_g},~~~r_g=+0,
\end{equation}
and replace Eq.(\ref{eq:bons0}) with  
\begin{equation}\label{eq:bonm}
\begin{split}
\left( \frac{\dot{M}}{4\pi}\right) ^2(r^3\rho^2)^{-1}+3(r^3\rho^2)^{1/3}=\\
\frac{r}{r-r_g}+(3+V^2)r.
\end{split}
\end{equation}

Again following Bondi, minimize the l.h.s. of Eq.(\ref{eq:bonm}) over $r^3\rho^2$:
\begin{equation}\label{eq:mlhs}
{\rm min}_{\rm ~lhs}=4\left( \frac{\dot{M}}{4\pi}\right)^{1/2}.
\end{equation}
Minimize the r.h.s. of Eq.(\ref{eq:bonm}) over $r$:
\begin{equation}\label{eq:mrhs}
{\rm min}_{\rm ~rhs}=1.
\end{equation}
Require that the minima coincide:
\begin{equation}\label{eq:bord}
\dot{M}=\frac{\pi}{4},
\end{equation}
which is the spherical Bondi rate in the dimensionless units (\ref{eq:uni}). 

To summarize in words: close to the accretor, the flow becomes spherically symmetrical, with the mass accretion rate given by the spherical Bondi formula, because the critical (sonic) surface of the spherical Bondi flow (with $\gamma=5/3$) is at $r=+0$. In fact, the sonic surface of a stationary axisymmetric flow with $\gamma=5/3$ touches the accretor at any Mach number \cite{Fogl}.

~

\section{Numerical Calculation}\label{sec:num} 

To solve the variational problem, Eqs.(\ref{eq:varp}, \ref{eq:act}, \ref{eq:lag}), we minimize the discretized action (\ref{eq:act}). With a logarithmic grid in $r$, we can accurately calculate the flow onto a small accretor in a large domain.

The elliptic PDE (\ref{eq:cont}) requires two boundary conditions. We have used the ``wind'' boundary condition at the outer sphere of radius $r=r_{\rm max}=10$ and the ``radial'' boundary condition at the inner sphere $r=r_{\rm acc}=0.01$:
\begin{equation}\label{eq:bc}
\phi(r_{\rm acc},\theta)=\phi_0,~~~\phi(r_{\rm max},\theta)=-Vr_{\rm max}\cos\theta .
\end{equation}

Then the constant $\phi_0$ is tuned to maximize the accretion rate. For $V=0.5$ (Mach number 0.5), we get the numerically computed accretion rate 
\begin{equation}\label{eq:nar}
\dot{M}(V=0.5)_{\rm num}=0.8344,
\end{equation}
in a 0.15\% agreement with the spherical Bondi rate (modified to account for a finite accretor radius) 
\begin{equation}\label{eq:tar}
\dot{M}_{\rm mod. Bondi}=\frac{\pi (1+3r_{\rm acc})^2}{4}=0.8332.
\end{equation}

The $\phi$-isolines of the flow are shown in Fig.(1). 

\begin{figure}
\includegraphics[width=0.5\textwidth]{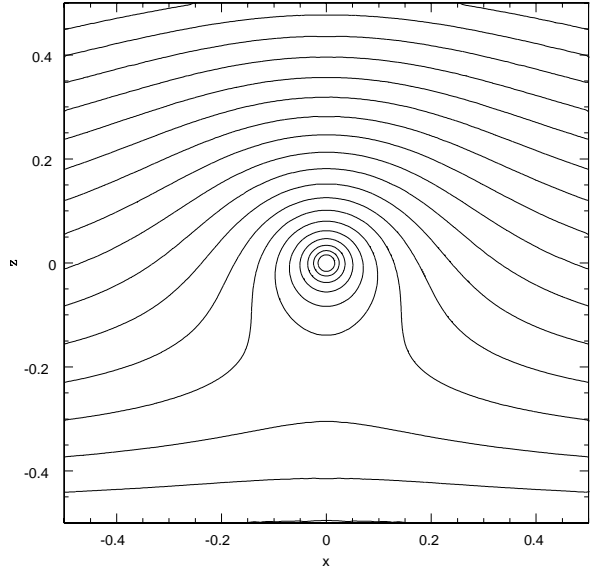}
\caption{Isolines of the velocity potential. Mach number 0.5. The spatial distances are in units of $R_{\rm B}\equiv \frac{2GM}{c_\infty^2}$. The accretor radius is $0.01R_B$. The outer boundary is at $10R_B$.}
\end{figure}

~

I thank Yasine Ali-Haimoud for many useful discussions.

~

\bibliographystyle{hapj}

\end{document}